\documentclass[runningheads]{llncs}
\usepackage{graphicx}
\usepackage{amssymb}
\usepackage{graphicx}
\usepackage{multirow}

\usepackage{booktabs}
\usepackage{soul}
\usepackage{comment}

\usepackage{ragged2e}  
\usepackage{array}
\usepackage{xspace}
\usepackage[colorinlistoftodos,prependcaption,textsize=footnotesize]{todonotes} 
\usepackage{pifont} 
\usepackage{epstopdf}
\usepackage{subfigure}

\usepackage[subtle]{savetrees}

\newcommand*{\thead}[1]{\multicolumn{1}{c}{\bfseries #1}}

\newcommand{\cmark}{\ding{51}}%
\newcommand{\xmark}{\ding{55}}%

\usepackage{url}
\usepackage{tablefootnote}
\usepackage{xcolor}

\begin{document}
\title{Towards Low-level Cryptographic\\ Primitives for JavaCards}

\author{Vasilios Mavroudis\inst{1} \and
Petr Svenda\inst{2}}
\authorrunning{V. Mavroudis and P. Svenda}
%
\institute{University College London\\
\email{v.mavroudis@ucl.ac.uk}\\
\and Masaryk University\\
\email{svenda@fi.muni.cz}}

\maketitle              
\begin{abstract}
JavaCard is a multi-application security platform deployed to over twenty billion smartcards, used in applications ranging from secure payments to telecommunications. While the platform is a popular choice for established commercial use cases (e.g., SIM cards in telecommunication networks), it has notably low adoption rates in: 1) application scenarios requiring recently-standardized cryptographic algorithms, 2) research projects, and 3) open source initiatives. We attribute this to the
restricted access to low-level cryptographic primitives (e.g., elliptic curve operations) and the lack of essential data types (e.g., Integers). While the underlying hardware has those capabilities, the JavaCard API does not provide calls for the corresponding functionality. Until now, the only available workaround was manufacturer-specific proprietary APIs that come with very restrictive non-disclosure agreements.

In this paper, we introduce a methodology to efficiently derive essential data types and low-level cryptographic primitives
from high-level operations. Our techniques are ideal for resource-constrained platforms, and make optimal use of the
underlying hardware, while having a small memory footprint. We also introduce JCMathLib, which, to the best of our knowledge, is the first generic library for low-level cryptographic operations in JavaCards that does not rely on a proprietary API. Without any disclosure limitations, JCMathLib enables open code sharing, release of research prototypes and
public and third-party code audits.

\end{abstract}

\vspace{-0.2cm}
\section{Introduction}\label{sect:introduction}\vspace{-0.2cm}
Smartcards were a breakthrough in the secure hardware world as they provided a relatively inexpensive platform that stores and computes over secrets securely while operating under adversarial control. At first, each hardware manufacturer maintained its platform that applications were compiled against. However, the need for cross-manufacturer interoperability eventually led to the formation of JavaCard platform~\cite{JCsdk}. JavaCard is a multi-application platform that promises full application portability between manufacturers. This fueled the adoption of the platform, and its popularity steadily rose over the years; until 2017 more than 20 billions of JavaCards have been sold~\cite{b_2017}. Common applications include telecommunications (e.g., SIM cards), payments, and loyalty cards. Additionally, new use cases emerged as Internet-of-Things (IoT) vendors seek secure hardware components for their devices~\cite{urien2017towards,kyrillidis2016smart,bichsel2009anonymous,diez2015toward}.

In practice, interoperability is achieved through the JavaCard Application Programming Interface (API), and the specification of the JavaCard runtime environment (JCRE) executing interoperable bytecode. The API is available in all JavaCards regardless of their manufacturer and is maintained by Oracle~\cite{JCsdk}, which periodically (approximately every three years) releases updates with JC 3.0.5 being the latest \cite{jc305}. JavaCard developers use the algorithms and the convenience functions it exposes in their applets, without any need to manage the underlying hardware (i.e., cryptographic coprocessor, secure parts of the memory). For instance, to use the Digital Signature Algorithm over Elliptic Curves (ECDSA), JavaCard developers can simply initialize a \textit{crypto engine} with \textit{ECDSA}. Behind the scenes, a hardware implementation of the algorithm is executed in the cryptographic coprocessor of the card, while all secrets are stored in a safe section of the memory. Such hardware-supported implementations of popular algorithms are very convenient for developers as they are much faster than their software-only counterparts and provide additional protection against side-channel and fault-induction attacks.

However, this reliance on the high-level API comes at a price. Developers do not have direct access to the underlying cryptographic components (e.g., crypto coprocessor) and rely solely on the API for the functionality they need. While the JavaCard API supports a range of common cryptographic algorithms, it cannot sufficiently cover the wealth of crypto schemes that are currently used in production. This results in many popular or modern algorithms not being available in JavaCards (e.g., El-Gamal~\cite{elgamal1985public}, Schnorr~\cite{schnorr1989efficient} or many Authenticated Encryption algorithms~\cite{bellare2000authenticated}) and thus hinders the adoption of JavaCards in novel applications.
To address these limitations, many manufacturers provide proprietary APIs that give some degree of freedom to developers to
implement unsupported schemes themselves. However, since those APIs are manufacturer-specific, their use severely hinders interoperability.
More importantly, the legal limitations proprietary APIs come with, make them unsuitable for many applications such as reproducible research, public code audits, application prototypes, and open source projects. In practice, proprietary APIs prohibit all applications that require any source code to be publicly released or shared.

This paper proposes a solution that goes beyond manufacturer solutions and is based solely on functionality provided by the standard API.
At a first glance, implementing new functionality in software may look straightforward. However, resource-constrained
platforms do not have neither the processing power, nor the memory capacity to support such implementations.
To overcome these limitations, we introduce a new approach that makes optimal use of the platform's resources (e.g., memory, hardware-acceleration chips).
We design various transformations that given a restricted set of high-level cryptographic operations,
can be used to realise essential low-level operations and fundamental data types. More specifically, our transformations
combine existing high-level hardware-accelerated operations (e.g., crypto-coprocessor for ECDSA), into new operations, data types and objects such as Integers, BigNumbers, EC Points, Modular arithmetic, Elliptic curve operations.
To evaluate the practicality of our approach and its performance, we build a library called \textit{JCMathLib}, which realizes several data types and primitives that are currently not supported by the JavaCard API. The JavaCard platform is an ideal testbed, as it has a very restrictive API and limited processing resources.

\medskip \noindent{\bf{Contributions}}\\
This paper makes the following contributions:

\begin{itemize}
\item \textbf{Transformations for primitive derivation.} We introduce a complete set of operation transformations that combine high-level
cryptographic operations to reconstruct low-level ones. Our transformations cover all fundamental cryptographic primitives, needed for
implementing any modern scheme, and span from simple Integer operations to resource-demanding elliptic curve points operations.

\item \textbf{Open-source cryptographic library.} \sloppy To the best of our knowledge, \mbox{\textit{JCMathLib}} is the first open source library for JavaCards that
realizes essential Integer, BigNumber, and primitive Elliptic Curve operations without the use of any proprietary APIs. This will
enable developers to publicly share parts of their code (e.g., research prototypes, code audits).

\item \textbf{Performance optimizations and evaluation.} The library's performance is extensively optimized and evaluated on various cryptographic smartcards.
The BigNumber operations are executed faster than reported in previous works \cite{tews2010ov}, while even the most computationally-intense EC Point operations
complete in 4 seconds or less. JCMathLib requires only 1KB of RAM in the performance-optimized version, while can also accommodate low-end cards with less RAM by moving part of its state from the RAM into the EEPROM.
\end{itemize}

The paper is organized as follows: We first provide some preliminary information on how
the JavaCard ecosystem operates and its shortcomings in the development and
adoption in new application areas. Then, we introduce a set of transformations
that realize commonly missing functionality in resource-constrained platforms,
and present JCMathLib, a library, that implements all these transformations for the
JavaCard platform. Finally, we evaluate and discuss the performance, security and
limitations of our library.

\vspace{-0.2cm}
\section{Preliminaries}\label{sect:prelim}\vspace{-0.2cm}
Smartcards are small embedded devices used in a wide range of applications that require secure storage, access control, and/or authentication. Initially, smartcards were cumbersome to program and manage as the functionality had to be loaded by the manufacturer at the time of the fabrication. This changed with multi-application cards supporting dynamic loading of multiple binaries. Building on top of this, JavaCard then introduced binary interoperability between different cards and manufacturers, thus increasing smartcard adoption even further. In this section, we first provide an overview of the JavaCard ecosystem processes, and then examine the JavaCard development and API.\\

\medskip \noindent{\bf{JavaCard Specification.}}
JavaCard holds the largest chunk of the programmable smartcard market with more than 20 billion units deployed in production~\cite{b_2017,mayes2014introduction}. To guarantee manufacturer interoperability, Oracle periodically releases the specifications of the JavaCard API (JC API), the JavaCard Virtual Machine (JCVM), and the JavaCard Runtime Environment (JCRE)~\cite{JCsdk}. The JCVM aims to isolate the user applications from the underlying hardware and operating system, while the JC API exposes JavaCard packages and classes to user applications. Manufacturers are responsible for implementing these specifications in their products and perform audits to ensure the security of the final product. The most recent version of the specification is JC $3.0.5$ (published in 2015), while the oldest specification available online is JC $2.1$ (published in 1999). Starting from JC 3.x, the specification distinguishes between two significantly different APIs -- the \emph{classic} and the \emph{connected} editions. The \emph{classic} edition further extends the capabilities of the JC 2.x versions but does not differ conceptually from previous versions. On the other hand, the \emph{connected} edition introduces smartcards exposing web-service interfaces accepting \textit{XML} requests. While the \emph{classic} edition is widely used, the \emph{connected} edition has yet to appear in commercial cards.

\medskip \noindent{\bf{JavaCard Development.}}
Even though JavaCard is a subset of Java, it comes with severe limitations that are mainly due to the low capabilities of the card's hardware. In particular, most of the standard Java packages have been stripped away and new packages have been added to support selected cryptographic operations and basic data transmission functions. Moreover, several standard Java features are missing (e.g., threads, dynamic class loading, and object cloning), while the JCVM specification further restricts the maximum number of classes to 15 and methods to 128 to make bytecode more compact.
From a technical perspective, JavaCard applet compilation is straightforward. Applets are compiled using ordinary Java tools (i.e., \emph{javac} compiler) configured to link the JavaCard packages. The produced Java bytecode is then processed and checked for compliance by a JavaCard converter and if successful, is turned into JavaCard bytecode. The final binary is then uploaded and installed to the card, usually via the GlobalPlatform interface~\cite{globalplatform}. To prevent potential security issues, JavaCard will refuse to load an applet that has been processed using a converter complying with a newer version of the JavaCard SDK. For instance, a smartcard shipped with JCVM 3.0.1 cannot load an applet converted using JCSDK 3.0.5, even if only packages and classes defined in 3.0.1 are used.\\

\noindent{\bf{JavaCard API.}}
JavaCards expose their functionality through various specialized packages that are part of the standard API.
In total, there are only 19 packages defined by the JC 3.0.5 API. Table~\ref{tab:api} lists
the packages that contain common high-level cryptographic operations,
while there are also specialized packages for biometric authentication and remote method invocation.
For performance reasons, the majority of those cryptographic algorithms are implemented in
specialized hardware components (i.e., dedicated crypto-coprocessors).

\begin{table}
\centering
\begin{tabular}{l  l }
\toprule
\thead{Package} & \thead{Functionality}\\
\hline \addlinespace
\emph{javacardx.crypto.Cipher} & Symmetric Encryption Schemes\\ \addlinespace
\emph{javacard.crypto.Signature} & Asymmetric Signing Schemes\\ \addlinespace
\emph{javacard.security.MessageDigest} & Cryptographic Hash Functions\\ \addlinespace
\emph{javacard.security.RandomData} & Random Number Generators\\ \addlinespace
\emph{javacard.security.KeyAgreement} & Key Agreement Algorithms\\ \addlinespace
\emph{javacard.security.KeyBuilder} & Key Generation\\ \addlinespace
\emph{javacard.security.KeyPair} & Key Pair Storage\\ \addlinespace

\bottomrule \addlinespace
\end{tabular}
\caption{List of JavaCard packages that provide cryptographic functionality.}
\label{tab:api}
\exhyphenpenalty 10000
\end{table}

JavaCard is notably slow in adopting modern schemes (or even variants of existing ones) and thus
developers are often faced with the challenge of implementing unsupported schemes by themselves.
Without access to the underlying hardware, developers can only build their schemes in software.
However, such implementations are rarely usable, as they are constrained both by the limited amount of RAM,
and the CPU's meager performance. For example, a hardware-accelerated
implementation of AES takes less than 2ms to execute, while a software
re-implementation requires more than 500ms~\footnote{\url{https://www.fi.muni.cz/~xsvenda/jcalgtest/comparative-table.html}}.
A solution to this problem could be the use of the \emph{javacardx.framework.math} package, specifying the \emph{BigNumber} class
with hardware-accelerated add and multiply operations.
However, manufacturers are not obligated to support all the packages and classes listed in the specification,
and \emph{BigNumber} is not supported by the great majority of the cards.
This lack of full specification compliance results in discrepancies in the functionality exposed by different cards.
For instance, an applet running fine in one card, will fail to install and instantiate properly on another card that does not support one
package, class or even an algorithm variant it uses (e.g., EC operations over large prime fields with 521-bit
length \emph{ALG\_EC\_FP, LENGTH\_EC\_FP\_521}). Currently, there is no complete list of smartcards and their
supported packages and developers rely primarily on trial and error, and the JCAlgTest survey~\cite{jcalgtest}.

On the other hand, manufacturers maintain their own proprietary APIs that provide solutions to many of those problems.
These APIs are vendor-specific and either provide functionality not exposed by the standard API, or
realize hardened and otherwise extended versions of the algorithms. For example, a commonly provided feature
is operations over elliptic curves, which are allow for the developement of fast implementations of unsupported
cryptographic algorithms. However, proprietary APIs are typically available to selected customers and only
after signing a non-disclosure agreement (NDA) that prevents the publication of any details, information and source code.

\vspace{-0.2cm}
\section{Low-level Primitive Derivation}\label{sect:transformations}\vspace{-0.2cm}
In this section, we introduce some generic transformations that can be used to efficiently derive low-level primitive operations from
high-level ones. We provide transformations for all the operations needed to implement any modern asymmetric cryptographic algorithm. Our approach assumes that a platform exposes a limited set of hardware-accelerated high-level operations (e.g., ECDSA signatures) and has some limited general computation capabilities. In the rest of this section, we focus on two types of low-level operations: modular arithmetic ones and operations on elliptic curves.

\medskip \noindent{\bf{Modular Arithmetic}}\\
Modulo operations are central in most cryptographic algorithm implementations. The main difficulty in implementing such operations in a resource-constrained platform is the computational load of handling and operating over large numbers. Moreover, the size of the numbers used in cryptography often exceeds the storage capacity of the native data types, and thus a specialized BigNumber type must be used. For simplicity, the rest of this section assumes that the platform exposes a BigNumber class. In Section~\ref{sect:jcmathlib}, we outline how such a class can be implemented in case the platform does not provide it natively.
\begin{table}
\centering
\begin{tabular}{l  c  c c}
\toprule
\thead{Primitive} & \thead{CPU} & \thead{Acceleration} & \thead{Operations}\\
\hline \addlinespace
Addition & \cmark & \xmark & - \\ \addlinespace
Subtraction & \cmark & \xmark & -\\ \addlinespace
Negation & \cmark & \xmark & -\\ \addlinespace
Exponentiation & \xmark & \cmark & RSA Decryption\\ \addlinespace
Multiplication & \cmark & \cmark & RSA Decryption\\ \addlinespace
Inversion & \cmark & \cmark & RSA Decryption\\ \addlinespace
EC Point Generation & \xmark & \cmark & EC Key Pair Generation\\ \addlinespace
EC Point Inversion & \cmark & \xmark & Modular Operations\\ \addlinespace
EC Point Addition & \cmark & \cmark & Modular Operations\\ \addlinespace
Scalar-EC Point Multiplication & \xmark & \cmark & EC Diffie-Hellman\\ \addlinespace
\bottomrule \addlinespace
\end{tabular}
\caption{List of all modular arithmetic and elliptic curve (EC) primitive operations realised by our transformations, their CPU and hardware-acceleration usage, and the operations relied on.}
 \label{tab:operations}
 \exhyphenpenalty 10000
 \end{table}

As seen in Table~\ref{tab:operations}, simple operations (i.e., \textit{addition}, \textit{subtraction}, \textit{negation}) are not too computationally
intensive and can be handled by the CPU without the use of any acceleration. On the other hand, \textit{multiplication},
\textit{exponentiation} and \textit{inversion} require the use of hardware-acceleration. To realise those operations, we
use the ``RSA Decryption'' routine, which is found in most of the platforms that support cryptography and is almost always hardware-accelerated.
More specifically, we use the plaintext version of RSA
to tunnel computations to the
coprocessor, and thus offload the CPU. For \textit{Modular multiplication}, the transformation is as straightforward as
simply executing an RSA Decryption. The RSA private key parameter holds the exponent operand, while the ciphertext and
the prime field parameter ones hold the base and the modulo operands respectively. As seen in Table~\ref{tab:operations},
no additional computations are performed on the CPU. A similar approach was also used in~\cite{bichsel2009anonymous}.

\textit{Modular multiplication} can be realized in a similar manner. However, the transformation makes use of both the CPU and the coprocessor as there is no commonly used cryptographic scheme that directly triggers a hardware-accelerated multiplication. Instead, as seen in Formula~\ref{eq:mult}, we express multiplication as a series of operations that are either hardware-accelerated or lightweight-enough for the CPU to execute:%
\begin{equation}\label{eq:mult}
x\cdot y\ mod\ p\ =\ ((x+y)^2-x^2-y^2)/2\ mod\ p
\end{equation}
From these operations, the modular exponentiations are performed on the
coprocessor, while the subtractions, addition and the division by two (i.e., bit shift) are handled by the CPU.
While \textit{Modular Inversion} may look like a special case of exponentiation, in many platforms the private key (i.e., exponent)
cannot be a negative number. Hence, RSA Decryption could not be directly used to perform the operation. However, given that our transformations are geared towards cryptographic applications, we optimize ``inversion'' for use with prime moduli.
In particular, based on prime field properties we express the inversion
as seen in Equation~\ref{eq:inv}:
\begin{equation}\label{eq:inv}
x^{-1}\ mod\ p\ =\ x^{p-2}\ mod\ p
\end{equation}
Hence, at the cost of some generality, we offload the CPU completely. A software-based solution can still be available
for cases where $p$ is not a prime. Finally, we realize a more specialized low-level operation that solves for $r$, a congruence
of the form $r^2\ =\ n\ (mod\ p)$, where $p$ is a prime. This operation is commonly used when decompressing the coordinates of
elliptic curve points. For that, we use the Tonelli Shanks algorithm~\cite{shanks1972five}~\cite{tonelli1891bemerkung}, and the previously described modular multiplication and exponentiation operations as building blocks.

\medskip \noindent{\bf{Elliptic Curve Arithmetic}}\\

Operations over elliptic curves are also essential for most modern cryptographic algorithms.
We provide transformations for four elliptic curve operations: \textit{EC Point Generation}, \textit{EC Point Inversion}, \textit{EC Point Addition} and \textit{Scalar-EC Point Multiplication}.
Similarly, with above, we also assume a data type capable of storing elliptic curve (EC) points, while in Section~\ref{sect:jcmathlib}, we explain how such a structure can be realised, when it is not provided natively by the platform.

For the generation of EC Points, we rely on the platform's Key Pair generation capabilities. The on-chip EC Key Pair generation uses a hardware random number generator. The public part of these keys is essentially a random point on the curve. Alternatively, given that there is a reliable source of randomness generating random EC points is trivial. The inverse of an EC point $P = (x, y)$ is $P^{-1}=(x, -y)$, which is also easy to compute. We simply use the modular negation operation derived previously, to negate the $y$ coordinate.
Point addition is more complex and requires a combination of hardware-accelerated operations and computations on the CPU.
We choose to use the ``double \& add'' algorithm and re-use the modular arithmetic operations outlined above. However,
given that the modular arithmetic operations have been realised, any other EC point addition algorithm can also be used.
Finally, the multiplication of a scalar $s$ with a point $P$ can be realized using a hardware-accelerated version
of the EC Diffie-Hellman (ECDH) Key Agreement protocol. This protocol is commonly implemented in resource-constrained platforms

and works as follows:
Alice and Bob want to agree on a common secret. At first, they generate their own EC key pairs
($a,A)$ and $(b,B)$ and then share their public keys $A$ and $B$ respectively. Alice uses
the ECDH routine in her platform to compute $a \cdot B$, while Bob computes $b \cdot A$ in the same manner.
The result of this operation is their shared secret. In our transformation, we abuse this
ECDH call to multiply any arbitrary $s$ with a $P$ of our choice, and get the result as the ``agreed secret''.

\vspace{-0.2cm}
\section{JCMathLib}\label{sect:jcmathlib}\vspace{-0.2cm}
We now introduce \textit{JCMathLib}, a cryptographic library for the JavaCard platform.
JCMathLib extends the standard JavaCard API by realizing various data types, primitive operations and convenience methods that are currently either not specified or not available.
By providing these operations and data types, JCMathLib enables developers to
implement cryptographic algorithms not supported by the JavaCard specification. As such, the library focuses not only on provisioning a complete set of low-level cryptographic primitives
necessary for the implementation of many cryptographic algorithms but also on some additional convenience classes.
We first provide an overview of the library's design and then discuss integration aspects and strategies
used to overcome the platform's limitations.

\subsection{Structure}\label{sect:design}\vspace{-0.2cm}
JCMathLib provides four classes that realize functionality applicable to different use cases.
Some of those classes are outlined in the API specification but are not available in commercial JavaCards, while others are completely new.

\medskip\noindent{\textit{Integer class.}}
The primitive data type \textit{int} and its wrapper class \textit{Integer} are listed in the JC virtual machine specification since 2002~\cite{vmspecification22} as ``optional''.
We implemented the class according to the specification to store a 32-bit signed value along with methods for addition, subtraction, multiplication, division and modulo operations.

\medskip \noindent{\textit{BigNumber class.}} This class encapsulates a large unsigned value of arbitrary length and provides methods for performing arithmetic operations on it. 
The BigNumber class both complies and extends the API specification. It has flexible internal representation capacity~\cite{specification222} and provides methods for: \textit{addition}, \textit{subtraction}, \textit{multiplication}, \textit{division}, \textit{modulo}, \textit{exponentiation}  and their modular counterparts for 8-bytes and longer numbers.

\medskip \noindent{\textit{ECCurve class.}} Stores the parameters of a given elliptic curve and provides a convenient way to handle multiple elliptic curves in a single applet. The class provides calls for key pair generation over a specific curve and is a dependency of the ECPoint class.

\medskip \noindent{\textit{ECPoint class.}} This class implements an Elliptic Curve Point and provides methods for \textit{point addition}, \textit{inversion}, \textit{doubling} and \textit{scalar multiplication}. Similar classes can only be found in proprietary APIs, as the JC API specification does not define it. This class in conjunction with BigNumber enables the implementation of
currently unsupported cryptographic algorithms based on ECC. This class has a dependency on the ECCurve class.

\subsection{Implementation}\label{sect:impl}\vspace{-0.2cm}
JCMathLib is comprised of several source code files that implement the classes outlined above, as well as other complementary methods.
It can be loaded into any JavaCard project as an external package. However, it is advisable that
the developer strips away any unneeded functionality to reduce the size of the final applet.

\medskip \noindent{\bf{Technical Details.}}
As seen in Table~\ref{tab:operations}, most of the modular arithmetic operations rely on the RSA decryption algorithm.
To exploit the RSA decryption method, we first create a \textit{PrivateKey} object and set its value to be equal to that of the exponent operand.
Then, we initialize a cipher engine for RSA decryption (i.e., ALG\_RSA\_NOPAD) and pass the base and modulus operands as its parameters.
To store the resulting numbers, we implemented the BigNumber class using arrays of \textit{short} variables (i.e., the largest data type natively supported by JavaCard). Similarly, to store the coordinates of EC Points, we abuse the \textit{ECPublicKey} class, which is designed to hold the public part of an EC key pair. To implement the elliptic curve operations, we mainly rely on the modular arithmetic operations derived above, except the multiplication of a scalar $s$ with
an EC point $P$. There, we also invoke ALG\_EC\_SVDP\_DH\_PLAIN that realizes the EC Diffie-Hellman Key Agreement protocol. However, due to the design of the method, its output in JavaCard provides only the $x$ coordinate of the resulting EC point (the $y$ is not returned\footnote{Compressed elliptic curve points contain the $x$ coordinate in full and the first bit of coordinate $y$. During decompression, the two candidates $y$'s are computed, and the one that matches the bit is chosen.}). To obtain the missing $y$, we use $x$ and solve the Weierstrass equation for $y$. This gives us two candidate points $A=\langle x, y \rangle$ and $B=\langle x, \hat{y} \rangle$, where $y$ is complementary to $\hat{y}$.
To determine which point is the correct one, we extend the transformation to use another hardware-accelerated method: ECDSA~\cite{johnson2001elliptic}.
Using ECDSA, we first sign a dummy plaintext using $s$ as the private key and then verify the produced signature using $A$ as its public counterpart.
If the verification is successful, then $A$ is the result of the multiplication. Otherwise, it is $B$.
It should be noted that version JC 3.0.5 introduces a new method ALG\_EC\_SVDP\_DH\_PLAIN\_XY that returns both the $x$ and $y$ coordinates.
Once commercial cards supporting this version become available, our implementation can be shortened to a single API call resulting in significantly shorter execution time.

\medskip \noindent{\bf{Optimizations.}}

\begin{figure}
    \centering
    \includegraphics[trim = -1mm 0mm -1mm 0mm, clip, width=0.5\textwidth]{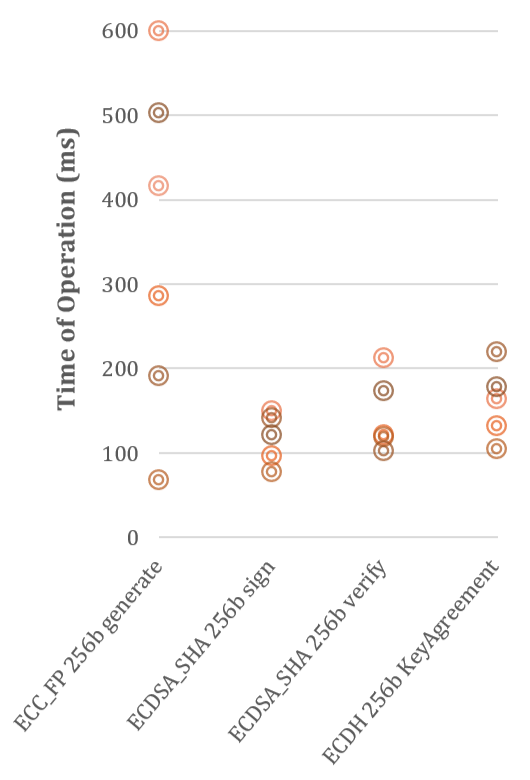}
    \caption{The measurement of basic operations available via JavaCard public API on six different smartcards.
    Every point corresponds to a single operation executed on a particular smartcard. The JCMathLib operations are then composed of such operations.}
    \label{fig:schema}
\end{figure}

As outlined in Section~\ref{sect:transformations}, our transformations rely on existing high-level operations
exposed by the platform. Naturally, this imposes an upper bound on speed, as
optimizations can reduce the computational load on the slow CPU,
but will still rely on the cryptographic coprocessor's performance.
Figure~\ref{fig:schema} illustrates the timings of certain high-level operations,
while it also highlights the timing differences between different cards.
Apart from the speed optimizations, memory management in such a constrained environment is also cumbersome.
Due to the complexity of the cryptographic operations, JCMathLib needs to use a large number of temporary variables to store intermediate results.
However, in contrast to standard Java programming patterns, in JavaCard, frequent object allocation introduces an impractically large performance overhead.
Pre-allocating all the temporary objects is not an option either, as the size of the fast transient memory is very limited. We, therefore,
developed a method to allocate and control a pool of shared memory objects residing in both the transient and the persistent memory.
Through the \textit{lock()} and \textit{unlock()} methods, the developer assigns objects to temporary variables, while our runtime logical checks
ensure that every object is used by only one temporary variable at a time. Optionally, the contents of a shared object can be automatically erased
upon locking or releasing the object to prevent information leakage.
This centralized object management system also allows for optimal object placement. A developer optimizing for speed can store the objects in the
faster transient memory, while a reduced transient memory footprint can be achieved through placement in the slower persistent memory.
These optimizations significantly improved the initial performance of JCMathLib and enabled us to support a wider range of cards, including some with
a limited amount of transient memory.

\medskip \noindent{\bf{JavaCard Performance Profiler.}}
Performance profiling of JavaCard code in applets is a notoriously difficult task. More specifically, as the card environment is build to protect the stored and processed secrets against an attacker with direct physical access, it is difficult to obtain a precise timing trace of the executed code on the granularity of separate methods or even lines of code.

To the best of our knowledge, there is no open-source performance profiler available for the JavaCard platform. As the optimizations of JCMathLib required an extensive analysis, we build a custom performance profiler called \textit{JCProfiler}. 
The profiler is based on the following idea: The source code of an applet is automatically extended with numerous additional lines of code called ``performance traps''. These traps are capable of prematurely interrupting the applet's execution when the stop condition matches their \textit{trap ID}. Such a \textit{trap} is inserted to parts relevant for measurement or after every single line in the applet's original code for the finest profiling granularity. Then, the client-side testing application repeatedly executes the applet for all possible controlling \textit{trap IDs}. As a result, an increasingly larger chunk of the applet's code is covered before the execution is interrupted by the corresponding \textit{trap}. The time measurements are collected and processed to estimate the time difference between the two consecutive traps -- resulting in the time required to execute a block of an original code between these two traps.
The intended position of a performance trap is defined by the developer who manually inserts a \textit{trap placeholder} in the intended line. \textit{JCProfiler} then locates these placeholders, inserts the conditional code with a unique \textit{trap ID} and generates the necessary client-side code to trigger all of them. The modified applet is then compiled and uploaded to the card. After the profiling session, the measured timings (in milliseconds) are automatically inserted in the applet source as comments next to the relevant lines.
For the profiling to be accurate two pre-conditions must be met: 1) the measured code needs to be re-entrant and deterministic (i.e., subsequent executions have the same behavior) and 2) the time required to reach a particular trap must fluctuate only modestly when repeatedly executed.  The reported timings carry an error that adds up over as more code is covered if these conditions are not satisfied.

\subsection{Evaluation}\label{sect:eval}\vspace{-0.2cm}
In this section, we study the performance and security provided by JCMathLib, while we also discuss other platform and licensing limitations.\\

\begin{table}
\centering
\begin{tabular}{l  c  c  c}
\toprule
 & \textbf{NXP J2E081} & \textbf{NXP J2D081} & \textbf{G\&D S@C6.0}\\
\hline \addlinespace
\textbf{BigNumber operations} & & \\
\hline \addlinespace
add(256b) & 7 ms & 10 ms & 10 ms\\ \addlinespace
subtract(256b) & 14 ms & 22 ms & 11 ms\\ \addlinespace
multiplication(256b) & 112 ms & 113 ms & 117 ms\\ \addlinespace
mod(256b) & 30 ms & 31 ms & 23 ms\\ \addlinespace
mod\_add(256b, 256b) & 71 ms & 72 ms & 56 ms\\ \addlinespace
mod\_mult(256b, 256b) & 872 ms & 855 ms & 921 ms\\ \addlinespace
mod\_exp(2, 256b) & 766 ms & 697 ms & 667 ms\\ \addlinespace
\hline \addlinespace
\textbf{ECPoint operations} & & \\
\hline \addlinespace
randomize(256b) & 296 ms & 245 ms & 503 ms\\ \addlinespace
add(256b) & 2995 ms & 2892 ms & 2747 ms\\ \addlinespace
inversion(256b) & 112 ms & 109 ms & 94 ms\\ \addlinespace
multiplication(256b)& 4157 ms & 3981 ms & 3854 ms\\ \addlinespace
\hline \addlinespace
\textbf{RAM overhead} & & \\
\hline \addlinespace
JCMathLib(256b) & 1144 B & 1152 B & 923 B\\ \addlinespace
new ECPoint(256b) & 0* B & 0* B & 0* B\\ \addlinespace
new BigNumber(256b)& 32 B & 32 B & 32 B\\ \addlinespace

\bottomrule \addlinespace
\end{tabular}
\caption{JCMathLib operations runtime(ms) and transient RAM overhead (Bytes) on three different commercials JavaCards. *Counter-intuitively the ECPoint class has no RAM overhead. This is because the ECPoint class wraps the KeyPair class which is always stored in persistent EEPROM.
}
\label{tab:jcmathlib_perfs}
\exhyphenpenalty 10000
\end{table}

\noindent\textbf{Performance}
Table~\ref{tab:jcmathlib_perfs} shows the runtime and the RAM overhead of the core JCMathLib operations and structures. We first observe that while there are some timing and RAM overhead discrepancies between
different cards, the measurements are quite consistent. Moreover, all the BigNumber operations
have a very low RAM overhead (32 bytes), while their runtimes are always smaller than 1000ms.
On the other hand, some of the ECPoint methods exhibit a higher runtime.
This is primarily due to the complexity of the transformations outlined in Sections~\ref{sect:transformations} and~\ref{sect:impl}. From these, the most expensive operations are the \textit{scalar multiplication}, which takes approximately 4s in some cards and the \textit{point addition}, which in our experiments took up to 4s as well.
To put these timings into context, according to the JCAlgTest performance measurements ~\cite{jcalgtest}
there are methods in the standard API with runtimes longer than 9s (e.g., RSA2048 CRT generate).
We also expect the multiplication to be significantly speed-up with the introduction of ALG\_EC\_SVDP\_DH\_PLAIN\_XY as outlined in Section~\ref{sect:impl}.
Overall, while our library's primary goal is to enable code sharing and research and prototyping, its performance is decent.\\

\noindent\textbf{Security}
Hardware implementations are inherently more resilient to physical attacks than software ones. As such, APIs that are implemented purely in hardware are more robust to power analysis compared to JCMathLib.
On the other hand, proprietary APIs often prevent developers from releasing their code for public security auditing, as many proprietary API NDAs prohibit any form of public code sharing. JCMathLib allows developers to replace the proprietary API calls, perform dynamic analysis (test vectors, fuzzing, etc.) and publish their source code for review.\\
\\\vspace{-0.1cm}

\noindent\textbf{Limitations \& Licensing}
JCMathLib is based on the JavaCard 3.0 version with most modern cards with ECC support expected to be compatible. Moreover, the \textit{ECPublicKey} class and the \textit{KeyAgreement} engine support elliptic curves parameters expressed in Weierstrass form only. In practice, this is not expected to significantly affect the applicability of JCMathLib as i) the great majority of commonly used curves use this form~\cite{sec20002}~\cite{lochter2010elliptic}, and ii) curves in other forms can often be expressed in Weierstrass form\footnote{\url{https://github.com/david-oswald/jc_curve25519}}. Finally, smartcard RSA engine must accept the public exponent equal to 2 for the computation to efficiently compute squaring for large numbers. Apart from these requirements, the library does not come with any other restrictions and is released under the MIT open source license.

\section{Related Work}\label{sect:relatedwork}\vspace{-0.2cm}
Despite the limitations of the platform, JavaCards have been utilized in various applications both in research and production environments. Many of these works use smartcards as a secure element, while others aim to scrutinize the security features of the platform.

\medskip \noindent{\bf{Applications.}}
Many other works utilize the security features of the platform to realize new algorithms enabling new use cases and applications. The authors in~\cite{pujol2011ttp} use JavaCards to store election secrets in a distributed manner. For this purpose, they developed an applet realizing ElGamal encryption and Shamir's secret sharing scheme. However, as the JavaCard API provides no support for these algorithms, they had to abuse the API to realize the necessary functionality in a hybrid fashion. \cite{Mavroudis:2017:TEH:3133956.3133961} uses an array of JavaCards originating from different manufacturers to realize an architecture resilient to backdoors and hardware trojans. Their prototype features few hundreds of JavaCards loaded with an applet realizing distributed protocols for essential cryptographic operations (i.e., key generation, decryption, signing).  A software re-implementation of various cryptographic algorithms for the JavaCard platform is provided in \cite{svendaAES}, optimized for performance and low memory footprint. A follow-up work~\cite{kubandabc} benchmarks the speed of various basic cryptographic operations on different smartcards. In other works, JavaCard is used as an embedded secure element to make the device self-authenticable and add near-field communication (NFC) capabilities~\cite{diez2015toward}. Furthermore, in~\cite{bichsel2009anonymous} Bichsel et al. design and implement a lightweight anonymous credential system suitable for secure identity tokens. Even though they were using an older JavaCard version (2.2.1), they achieve transaction times that are orders of magnitudes faster than those of prior works. Other studies focus on more recent use cases such as securing bitcoin transactions/storage~\cite{urien2017towards} and homogenizing the Web of Things~\cite{kyrillidis2016smart}.
However, in some cases, the limitations of the JavaCard platform were impossible to overcome. For instance, Mostowski et al.~\cite{mostowski2011efficient} developed a lightweight anonymous credentials protocol suitable for devices with low computational power. Even though JavaCard hardware was sufficient for their protocols,
they opted to use MULTOS which is less widespread and without publicly available SDK.
Their main motivation was that MULTOS provides direct access to the underlying hardware and low-level cryptographic operations, while the JavaCard API does not.

\medskip \noindent{\bf{Third-party Libraries.}}
There is also a smaller category of works aiming to extend the JC API with low-level primitives. The most known project is OV-Chip 2.0~\cite{tews2010ov} which implement the BigNat library to provide big integer functionality for JavaCards. However, the library is developed with the OV-chipkaart in mind, and its maintenance has seized for more than eight years.
E-verification2~\cite{pujol2011ttp} featured a class that provides capabilities similar
to those of the Java BigInteger. Unfortunately, it was never released as a standalone library, and the development stopped in 2012. Finally, the JCMath library~\cite{JCMath} aimed to enable operations over large numbers in JavaCard, but it is neither maintained nor has any documentation available.
The underlying coprocessor for fast modular multiplication was utilized for fast implementation of quantum-computer resistant Kyber algorithm \cite{errorRSA} on smartcards; however, a low-level proprietary API was used and a working research prototype was not released.\\\vspace{-0.2cm}

\section{Conclusions}\label{sect:conclusion}\vspace{-0.2cm}
This paper identifies multiple issues that hinder the adoption
of the JavaCard platform in new use cases, and proposes techniques to alleviate them.
We first introduce a set of transformations that exploit existing high-level functionality
to reconstruct missing low-level primitive operations.
These transformations are geared towards resource-constrained platforms and
try to offload the CPU by utilizing any hardware-acceleration components that may be available.
We then use them to implement JCMathLib, a library, that realizes both
new features and already specified primitives that are not available in commercial
JavaCards. Our library provides developers with the necessary classes and methods
(e.g., BigNumber operations, Elliptic Curve Point operations)
to implement cryptographic algorithms not supported by the JC API.
This, for the first time, enables the rapid adoption of new algorithms, decreases the impact of the manufacturer compliance issues and makes the JavaCard platform a suitable candidate for previously unsupported applications.

\bibliographystyle{plain}
\label{sect:bib}
\bibliography{reference}

\begin{thebibliography}{10}

\bibitem{errorRSA}
Martin~R. Albrecht, Christian Hanser, Andrea Hoeller, Thomas Pöppelmann,
  Fernando Virdia, and Andreas Wallner.
\newblock Learning with errors on rsa co-processors, 2018.
\newblock \url{https://eprint.iacr.org/2018/425}.

\bibitem{bellare2000authenticated}
Mihir Bellare and Chanathip Namprempre.
\newblock Authenticated encryption: Relations among notions and analysis of the
  generic composition paradigm.
\newblock In {\em Proceedings of the ASIACRYPT 2000}, pages 531--545. Springer,
  2000.

\bibitem{bichsel2009anonymous}
Patrik Bichsel, Jan Camenisch, Thomas Gro{\ss}, and Victor Shoup.
\newblock Anonymous credentials on a standard {J}ava {C}ard.
\newblock In {\em Proceedings of the ACM CCS'09}, pages 600--610. ACM, 2009.

\bibitem{JCMath}
Alberto Carp.
\newblock Smartcard primitives repository.
\newblock
  \url{https://github.com/albertocarp/Primitives_SmartCard/blob/master/src/sid/JCMath.java},
  2016.

\bibitem{jcalgtest}
CRoCS-MUNI.
\newblock \uppercase{JCA}lg\uppercase{T}est -- {J}ava{C}ard algorithm support
  testing project.
\newblock \url{https://jcalgtest.org}, 2018.

\bibitem{diez2015toward}
Fidel~Paniagua Diez, Diego~Suarez Touceda, Jose Maria~Sierra Camara, and
  Sherali Zeadally.
\newblock Toward self-authenticable wearable devices.
\newblock {\em IEEE Wireless Communications}, 22(1):36--43, 2015.

\bibitem{elgamal1985public}
Taher ElGamal.
\newblock A public key cryptosystem and a signature scheme based on discrete
  logarithms.
\newblock {\em IEEE transactions on information theory}, 31(4):469--472, 1985.

\bibitem{globalplatform}
GlobalPlatform.
\newblock Global{P}latform {C}ard {S}pecification, 2018.
\newblock \url{https://globalplatform.org/specifications/technical-overview/}.

\bibitem{johnson2001elliptic}
Don Johnson, Alfred Menezes, and Scott Vanstone.
\newblock The elliptic curve digital signature algorithm ({ECDSA}).
\newblock {\em International Journal of Information Security}, 1(1):36--63,
  2001.

\bibitem{b_2017}
B~Karen.
\newblock The impact of {J}ava {C}ard {T}echnology yesterday and tomorrow.
\newblock
  \url{https://javacardforum.com/2017/03/02/the-impact-of-java-card-technology-yesterday-and-tomorrow/},
  2017.

\bibitem{kubandabc}
Karol Kubanda.
\newblock Comparison of the speed of cryptographic operations running on
  smartcards.
\newblock Master's thesis, Masaryk University, Czech Republic, 2008.

\bibitem{kyrillidis2016smart}
Lazaros Kyrillidis, Sheila Cobourne, Keith Mayes, and Konstantinos
  Markantonakis.
\newblock A smart card web server in the web of things.
\newblock In {\em Proceedings of SAI Intelligent Systems Conference}, pages
  769--784. Springer, 2016.

\bibitem{lochter2010elliptic}
Manfred Lochter and Johannes Merkle.
\newblock Rfc 5639: Elliptic curve cryptography ({ECC}) {B}rainpool standard
  curves and curve generation.
\newblock Technical report, 2010.

\bibitem{Mavroudis:2017:TEH:3133956.3133961}
Vasilios Mavroudis, Andrea Cerulli, Petr Svenda, Dan Cvrcek, Dusan Klinec, and
  George Danezis.
\newblock A touch of evil: High-assurance cryptographic hardware from untrusted
  components.
\newblock In {\em Proceedings of the ACM CCS'17}, pages 1583--1600, New York,
  NY, USA, 2017. ACM.

\bibitem{mayes2014introduction}
Keith Mayes and Konstantinos Markantonakis.
\newblock An introduction to smart cards and rfids.
\newblock In {\em Secure Smart Embedded Devices, Platforms and Applications},
  pages 3--25. Springer, 2014.

\bibitem{JCsdk}
\mbox{Oracle corp.}
\newblock Java {C}ard technology.
\newblock
  \url{http://www.oracle.com/technetwork/java/javame/javacard/index.html},
  2014.

\bibitem{jc305}
\mbox{Oracle corp.}
\newblock Java {C}ard {P}latform, {C}lassic {E}dition 3.0.5.
\newblock \url{https://docs.oracle.com/javacard/3.0.5/index.html}, 2015.

\bibitem{vmspecification22}
\mbox{Sun} Microsystems.
\newblock Java {C}ard 2.2 {V}irtual {M}achine {S}pecification.
\newblock
  \url{http://download.oracle.com/otndocs/jcp/7047-javacard_devkit-2.2_01-spec-oth-JSpec/},
  2002.

\bibitem{specification222}
\mbox{Sun} Microsystems.
\newblock Java {C}ard 2.2.2 {P}latform {S}pecification.
\newblock
  \url{http://download.oracle.com/otndocs/jcp/java_card_kit-2.2.2-fr-oth-JSpec/},
  2009.

\bibitem{mostowski2011efficient}
Wojciech Mostowski and Pim Vullers.
\newblock Efficient {U-P}rove implementation for anonymous credentials on smart
  cards.
\newblock In {\em International Conference on Security and Privacy in
  Communication Systems}, pages 243--260. Springer, 2011.

\bibitem{pujol2011ttp}
Jordi Pujol-Ahull{\'o}, Roger Jard{\'\i}-Ced{\'o}, Jordi Castella-Roca, and
  Oriol Farras.
\newblock {TTP} smartcard-based {E}l{G}amal cryptosystem using threshold scheme
  for electronic elections.
\newblock In {\em International Symposium on Foundations and Practice of
  Security}, pages 14--22. Springer, 2011.

\bibitem{schnorr1989efficient}
Claus-Peter Schnorr.
\newblock Efficient identification and signatures for smart cards.
\newblock In {\em Conference on the Theory and Application of Cryptology},
  pages 239--252. Springer, 1989.

\bibitem{sec20002}
SECG SEC.
\newblock 2: Recommended elliptic curve domain parameters.
\newblock {\em Standards for Efficient Cryptography Group, Certicom Corp},
  2000.

\bibitem{shanks1972five}
Daniel Shanks.
\newblock Five number-theoretic algorithms.
\newblock In {\em Proceedings of the second Manitoba conference on numerical
  mathematics}, volume~51, page~70, 1972.

\bibitem{tews2010ov}
Hendrik Tews.
\newblock Ov-chip 2.0 hacker's guide.
\newblock 2010.

\bibitem{tonelli1891bemerkung}
Alberto Tonelli.
\newblock Bemerkung {\"u}ber die aufl{\"o}sung quadratischer congruenzen.
\newblock {\em Nachrichten von der K{\"o}nigl. Gesellschaft der Wissenschaften
  und der Georg-Augusts-Universit{\"a}t zu G{\"o}ttingen}, 1891:344--346, 1891.

\bibitem{urien2017towards}
Pascal Urien.
\newblock Towards secure bitcoin fast trading: Designing secure elements for
  digital currency.
\newblock In {\em Mobile and Secure Services (MobiSecServ'17)}, pages 1--5.
  IEEE, 2017.

\bibitem{svendaAES}
Petr \v{S}venda.
\newblock Cryptographic algorithms re-implementation for {J}ava{C}ard.
\newblock \url{http://www.fi.muni.cz/~xsvenda/jcalgs.html}, 2014.

\end{thebibliography}

\end{document}